\documentclass[aps,pra,twocolumn,preprintnumbers,showpacs,superscriptaddress,amsmath,amssymb]{revtex4}
\usepackage{dcolumn}
\usepackage{bm}
\usepackage{amssymb}
\usepackage[german, english]{babel}
\usepackage{graphicx}
\usepackage{color}
\usepackage{textcomp}
\usepackage[letterpaper,total={7in,9.5in},top=0.75in,left=0.75in]{geometry}
\usepackage[utf8]{inputenc}
\usepackage{array}

\begin{document}

\title{Magneto-optical trapping of mercury at high phase space density}

\author{Quentin Lavigne}
\author{Thorsten Groh}
\author{Simon Stellmer}
\email{stellmer@uni-bonn.de}
\affiliation{Physikalisches Institut, Rheinische Friedrich-Wilhelms-Universität, 53115 Bonn, Germany}

\date{\today}

\begin{abstract}
We present a realization of a magneto-optical trap of mercury atoms on the $^{1}S_0 \rightarrow {}^3P_{1}$ intercombination line. We report on trapping of all stable mercury isotopes. We characterize the effect of laser detuning, laser intensity, and gradient field on the trapping performance of our system. The atom number for the most abundant isotope $^{202}{\rm Hg}$ is $5\times 10^7$ atoms. Moreover, we study the difference in cooling processes for bosonic and fermionic isotopes. We observe agreement with the Doppler cooling theory for the bosonic species and show sub-Doppler cooling for the fermionic species. We reach a phase space density of a few parts in $10^{-6}$, which consitutes a promising starting condition for dipole trap loading and evaporative cooling.
\end{abstract}

\maketitle

\section{Introduction}

All atoms in the class of alkaline-earth (-like) metal elements share a unique combination of properties:~two valence electrons and a $J=0$ ground state. The level structure decomposes into singlet and triplet states, the latter can be metastable and are connected to the single ground state via narrow intercombination lines. In recent years, this class of atoms has received wide-spread attention in the fields of optical clocks \cite{Ludlow2015}, in quantum simulation based on laser-cooled atoms \cite{Bloch2012}, and in low-energy searches for physics beyond the standard model \cite{Safronova2018,Cairncross2019}. Within this class of elements, mercury (Hg) assumes a unique role:~it is the heaviest element with stable isotopes that can be laser-cooled, it has the highest ionization threshold, and, as a consequence, all of its principal optical transitions are deep in the ultraviolet (UV) range. As the technology of UV lasers matured over the past decades, cold-atom experiments with mercury atoms became feasible.

Magneto-optical trapping of mercury has first been realized in seminal work by the group of H.~Katori \cite{Hachisu2008}, and forms the basis of optical clocks based on mercury \cite{Petersen2008,Mcferran2012,Mejri2011,Yamanaka2015,Ohmae2020}, benefiting in particular from its insensitivity to black-body radiation shifts. Related research on laser-cooling of mercury is also described in Refs.~\cite{Villwock2011,Mcferran2010,Paul2013,Liu2013,Witkowski2017}.

Here, we present a detailed study on laser-cooling of mercury. The identification of optimal parameter ranges, in combination with increased laser power, and has allowed us to substantially improve the atom number and phase space density of laser-cooled samples, compared to previous works. With these improvements, interesting experiments come into reach \cite{Fry1976}, including a competitive measurement of the Hg electric dipole moment (EDM) using laser-cooled atoms \cite{Graner2016,Parker2015}, isotope shift measurements \cite{Berengut2018,Witkowski2019,Berengut2020}, and evaporation towards degenerate quantum gases.

\section{Experimental setup}

In this work, laser cooling of neutral mercury is performed on the $^{1}S_0 \rightarrow {}^{3}P_{1}$ intercombination line at 254\,nm, which has a linewidth of $\Gamma = 2 \pi \times 1.3\,{\rm MHz}$ and a corresponding Doppler temperature $T_D = 31\,{\rm \mu K}$. The saturation intensity of this transition is $I_{\rm sat} = 10\,{\rm mW/cm}^2$. Note that precooling on the broad $^{1}S_{0} \rightarrow {}^{3}P_{0}$ singlet transition is challenging due to its wavelength of 185\,nm, where high-power cw laser development is still in its infancy \cite{Scholz2013}.

\begin{figure}
\includegraphics[width=\columnwidth, keepaspectratio]{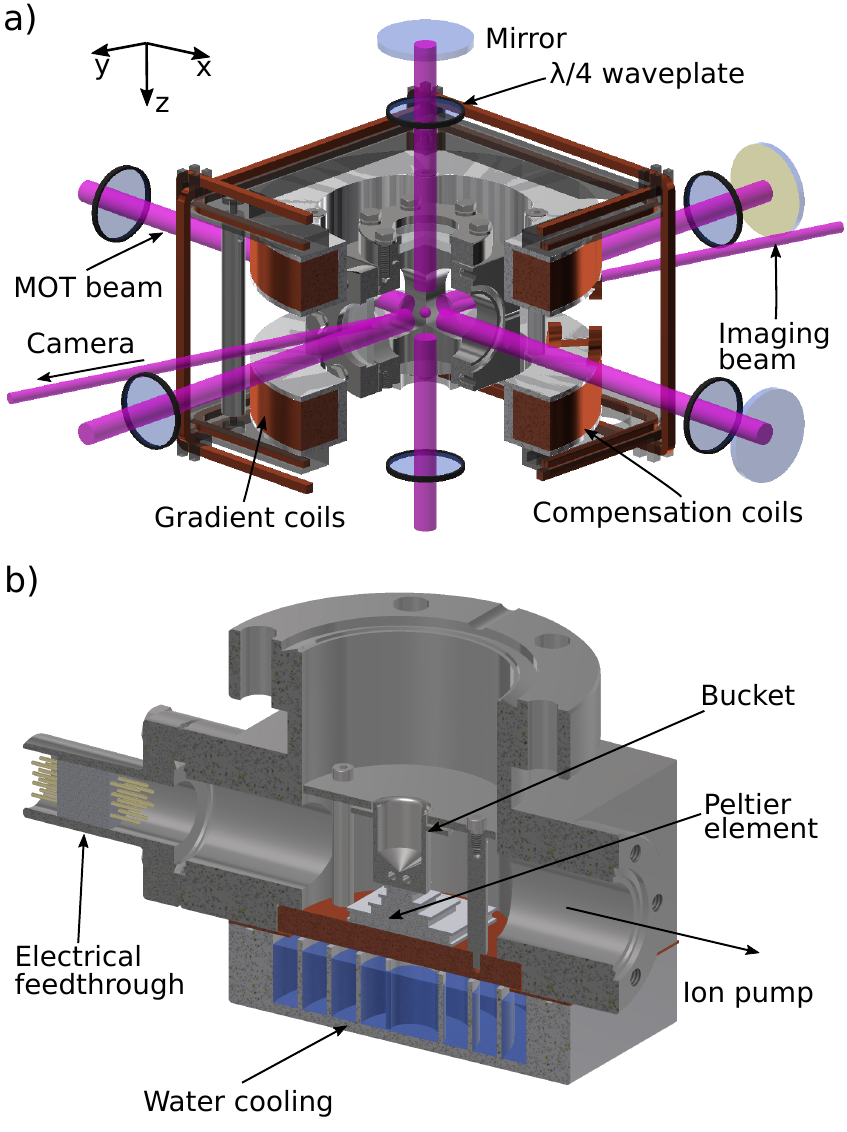}
\caption{Experimental setup. (a) Central region of the vacuum chamber, including coil systems and optics.  (b) Cut-away view of atomic source. A stainless steel bucket filled with liquid mercury is temperature-controlled via a four-stage Peltier element to set the vapour pressure in the vacuum chamber.}
\label{ExperimentalSetup}
\end{figure}

Our experimental apparatus, depicted in Fig.~\ref{ExperimentalSetup}, is designed as a test setup to identify optimal parameters for laser cooling of mercury. The mercury atoms are loaded from the background gas. The atom source is composed of a stainless steel reservoir filled with few droplets of liquid mercury. This reservoir is cooled under vacuum by a four-stage Peltier element down to $-50^{\circ}$C. For loading of a magneto-optical trap (MOT), the oven is operated at a temperature of $-35^{\circ}$C, resulting in a partial pressure of about $5\times10^{-6}\,$mbar at the source.  The source section is pumped with a 5\,L/s ion pump to protect the in-vacuum electronics from mercury corrosion. We observed recurrent failure of the 5\,L/s ion pump due to mercury poisoning after a power failure and subsequent increase of mercury partial pressure. Such a damage of the ion pump anode has also been observed in other studies \cite{Ward1966} and can be cured by high-potting or heating of the pump.

A CF40 tube with a length of 380\,mm (conductivity $\sim 1.55\,$L/s) connects the source to the MOT chamber. The vacuum chamber, which is assembled from standard CF40 vacuum components, is pumped down to the range of $10^{-8}\,{\rm mbar}$ by a 55\,L/s ion pump and a standard titanium sublimation pump.

The magnetic quadrupole field required for atom trapping is generated by a pair of coils in anti-Helmholtz configuration. The coils are made of 6\,mm $\times$ 6\,mm hollow-core square copper tubing and are water-cooled. They consist of 12 windings each and have a diameter of 160\,mm. These coils generate an axial gradient field ${\partial B_z}/{\partial z}$ of 0.20\,G/(cm A). In typical operation, the axial gradient field is set to about 10\,G/cm at a current of 50\,A. The low inductance of the coils ($\sim 100\,\mu$H) enables to quickly turn off ($< 1\,$ms) the magnetic field with an insulated-gate bipolar transistor. In practice, the switching time is limited by a metal frame to typically 6\,ms. Three mutually orthogonal pairs of coils in Helmholtz configuration allow to compensate Earth's magnetic field and any other background field.

The light at 254\,nm is generated from a commercial frequency quadrupled laser. The fundamental mode at 1016\,nm is generated by a diode laser, which is stabilized to a commercial high-finesse cavity (Menlo, finesse 74\,000 at 1016\,nm) for linewidth reduction. A fiber-coupled phase modulator (Jenoptik PM1064) with 7\,GHz bandwidth is used to imprint variable sidebands, which are used to steer the laser frequency with respect to the cavity mode.

The diode emission at 1016\,nm is amplified by a tapered semiconductor amplifier, passed through a filter of $\sigma = 5.7\,$GHz transmission bandwidth to remove undesired incoherent background radiation (ASE, amplified spontaneous emission of the semiconductor laser), and amplified by a fiber amplifier to about 8\,W. Two consecutive and resonant stages of second harmonic generation (SHG) generate more than 350\,mW of UV power.

Polarization components are used to split the light at 254\,nm into three pathways using: a spectroscopy cell for monitoring purposes, the MOT beams, and the imaging beam. The light for the MOT is passed through an acousto-optical modulator (AOM) for intensity and frequency control, before being split into three arms. The mean waist is increased up to $w_0 = 6.5\,$mm. The three mutually orthogonal MOT beams are retro-reflected and have a typical power of $P_{\rm MOT}=30\,$mW per beam. All beams are aligned with sub-mm precision to the center of the quadrupole field.

Absorption imaging is performed on the same optical transition. The imaging beam is passed through an AOM for frequency adjustment and switching, it is then focussed through a $100\,\mu$m pin hole for mode cleaning, expanded to a waist of 7.5\,mm, and delivered to the MOT region. It is linearly polarized and has a typical power of 1\,mW.  Our imaging system is composed of single lens in $2f$-$2f$ configuration to obtain a magnification of $M = 1$. A charged-coupled device (CCD) camera (ANDOR model iXon3 885) with quantum efficiency in excess of 30\% is used for imaging.

A typical measurement sequence consists of a 5\,s long MOT-loading phase in which the gradient field and the MOT beams are turned on. Then, the gradient field and MOT beam are switched off. Atom number and temperature of the atomic cloud are determined from standard time-of-flight (TOF) images.

\section{Results}

\subsection{Magneto-optical trapping of all seven stable isotopes}

\begin{table*}
  \caption{Naturally occuring mercury isotopes. For each isotope, we state the nuclear spin $I$, its natural abundance $\mathit{NA}$ \cite{LideHandbook}, the observed atom number normalized to the most abundant isotope $^{202}{\rm Hg}$, $N^\prime$, and the trapping efficiency $N^\prime/\mathit{NA}^\prime$. The latter is expressed as the ratio of normalized atom number and normalized natural abundance, and shows a strong deviation from unity only for the fermionic isotopes.
  \label{AbundanceTable}}
  \setlength\extrarowheight{2pt}
  \begin{ruledtabular}
    \begin{tabular}{c c c c c c c}
      \begin{tabular}{@{}c@{}} Isotope \end{tabular} &
      \begin{tabular}{@{}c@{}} Spin \\ statistics \end{tabular} &
      \begin{tabular}{@{}c@{}} Nuclear\\ spin \\ $I$ \end{tabular} &
      \begin{tabular}{@{}c@{}} Natural \\ abundance \\ $\mathit{NA}$ \end{tabular} &
      \begin{tabular}{@{}c@{}} Normalized \\ abundance \\ $\mathit{NA}^\prime$ \end{tabular} &
      \begin{tabular}{@{}c@{}} Normalized \\ atom number \\ $\mathit{N}^\prime$ \end{tabular} &
      \begin{tabular}{@{}c@{}} Trapping \\ efficiency \\ $N^\prime/\mathit{NA}^\prime$ \end{tabular} \\
      \hline
      $^{196}{\rm Hg}$ & bosonic & 0 & 0.15 & 0.0052 & 0.0043(13) & 0.83 \\
      $^{198}{\rm Hg}$ & bosonic & 0 & 9.97 & 0.3455 & 0.3220(12) & 0.93 \\
      $^{199}{\rm Hg}$ & fermionic & 1/2 & 16.87 & 0.5845 & 0.2166(10) & 0.37 \\
      $^{200}{\rm Hg}$ & bosonic & 0 & 23.10 & 0.8004 & 0.8200(12) & 1.03\\
      $^{201}{\rm Hg}$ & fermionic & 3/2 & 13.18 & 0.4567 & 0.0862(7) & 0.19\\
      $^{202}{\rm Hg}$ & bosonic & 0 & 29.86 & 1 & 1 & 1 \\
      $^{204}{\rm Hg}$ & bosonic & 0 & 6.87 & 0.2380 & 0.1970(9) & 0.83 \\
    \end{tabular}
  \end{ruledtabular}
\end{table*}

\begin{figure*}
\includegraphics[width=\textwidth, keepaspectratio]{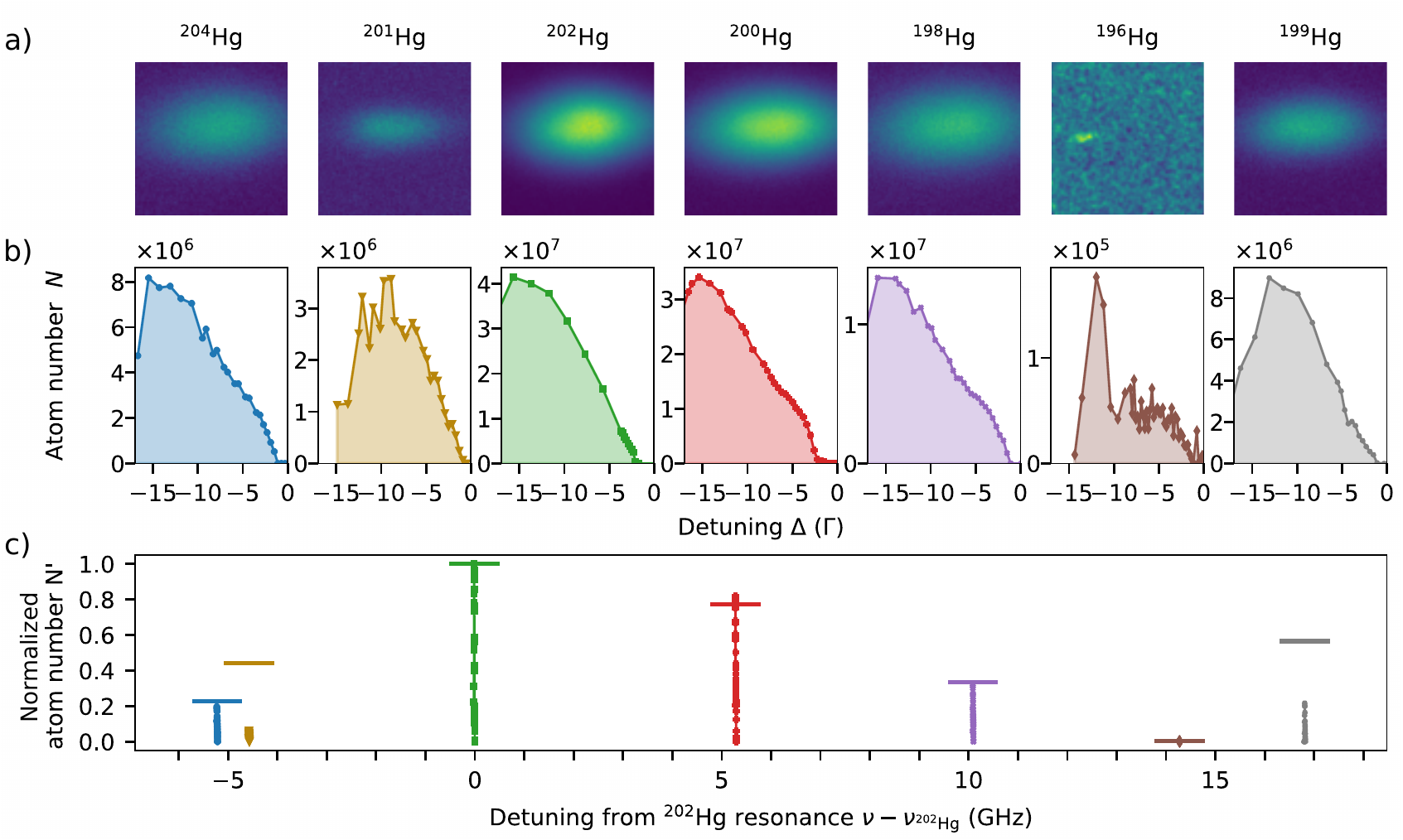}
\caption{Magneto-optical trapping of all stable Hg isotopes. (a) Absorption images of the MOTs after 0\,ms time of flight. The image of $^{196}{\rm Hg}$ is Gaussian filtered to increase visibility. (b) Number of trapped atoms $N$ as a function of the detuning $\Delta$ for each isotope. (c) Normalized atom number as a function of the detuning from the $^{202}{\rm Hg}$ resonance. The horizontal lines represent the normalized isotope abundance.}
\label{IsotopeTrapping}
\end{figure*}

We begin our study by presenting magneto-optical trapping of all seven stable mercury isotopes; see Fig.~\ref{IsotopeTrapping}. We adjust the waist of the MOT beams to $w_0 = 5.2\,$mm and set the intensity per beam to $I = 26\,$mW, corresponding to a saturation parameter $s = I / I_{\rm sat}$ of $s = 6$. The magnetic field gradient is set to ${\partial B_z}/{\partial z} = 10\,$G/cm. We scan the frequency of the MOT beams across the resonance frequency of each isotope. A maximum in atom number $N$ is reached for a detuning $\Delta$ of about $-15\,\Gamma$. We capitalize on the high laser power available, which allows us to increase the diameters of the MOT beams. For the first time, we were able to observe a MOT of the least abundant isotope, $^{196}{\rm Hg}$, with a natural abundance of only 0.15\%. This isotope has not been detected in previous studies \cite{Hachisu2008,Petersen2008,Villwock2011,Paul2013,Liu2013,Witkowski2017}.

A list of the stable isotopes of mercury is provided in Table \ref{AbundanceTable}:~there exist five bosonic isotopes with nuclear spin $I=0$ and two fermionic isotopes, $^{199}{\rm Hg}$ with $I=1/2$ and $^{201}{\rm Hg}$ with $I=3/2$. We observe that for the bosonic isotopes, the observed MOT atom numbers corresponds, within the uncertainties, to the natural abundances. For that we normalize the peak MOT atom numbers $N$ of each isotope to the one of the most abundant isotope, $^{202}{\rm Hg}$. We then compare the normalized MOT atom numbers $N^\prime = N / N_{202}$ of each isotope to its normalized natural abundance $\mathit{NA}^\prime = \mathit{NA} / \mathit{NA}_{202}$ (last column of table \ref{AbundanceTable}). The observed correspondence for the bosons is expected, as the electronic structure of these isotopes is exactly identical. This observation indicates that the MOT atom number is not yet saturated for the set of parameters used here.

For the fermions, however, we do observe a clear mismatch between normalized atom number and isotope abundance: cooling and trapping efficiency is reduced by a factor of about 3 for $^{199}{\rm Hg}$, and by a factor of about 5 for $^{201}{\rm Hg}$. Compared to the bosonic isotopes, these two isotopes possess multiple $m_F$ components in the ${}^1S_0$ ground state, as well as hyperfine and Zeeman structure in the ${}^3P_1$ excited state. The MOT is operated on the $F=1/2 \rightarrow  F^\prime=3/2$ transition for $^{199}{\rm Hg}$, and on the $F=3/2 \rightarrow F^\prime=5/2$ transition for $^{201}{\rm Hg}$. The reduced efficiency of fermionic MOTs has been explained in Ref.~\cite{Mukaiyama2003} and is observed with many alkaline-earth metal elements. In short, the vastly different $g$-factors of the ${}^1S_0$ ground state ($g\approx0$) and the ${}^3P_1$ excited state ($g \approx 1.5$), as well as the multitude of Zeeman states, reduce cooling power and open up loss channels.

While the vast majority of magneto-optical traps are operated on $F \rightarrow F^\prime=F+1$ transitions, there is also an interest to study unconventional MOT operation for the cases $F^\prime\leq F$. These cases are relevant for laser cooling of molecules and might use blue-detuned light \cite{Jarvis2018}. Indeed, we observe stable magneto-optical trapping of the $^{199}$Hg isotope on the $F=1/2 \rightarrow F^\prime=1/2$ transition. With similar trap parameters we reach around $N = 1.1(2) \times 10^4$ atoms at a detuning of $\Delta \approx -3\, \Gamma$. This is about two orders of magnitude reduction with respect to the ``ordinary'' $^{199}$Hg $F=1/2 \rightarrow F^\prime=3/2$ MOT.

In the following, we will focus our studies on the most abundant isotope $^{202}{\rm Hg}$. We will explore the key parameters such as laser detuning, intensity, and magnetic field gradient to optimize the performance of the experiment. These measurements significantly expand previous studies \cite{Mcferran2010,Mejri2011} to a broader parameter range.

\subsection{Atom number}

\begin{figure}
\includegraphics[width=\columnwidth, keepaspectratio]{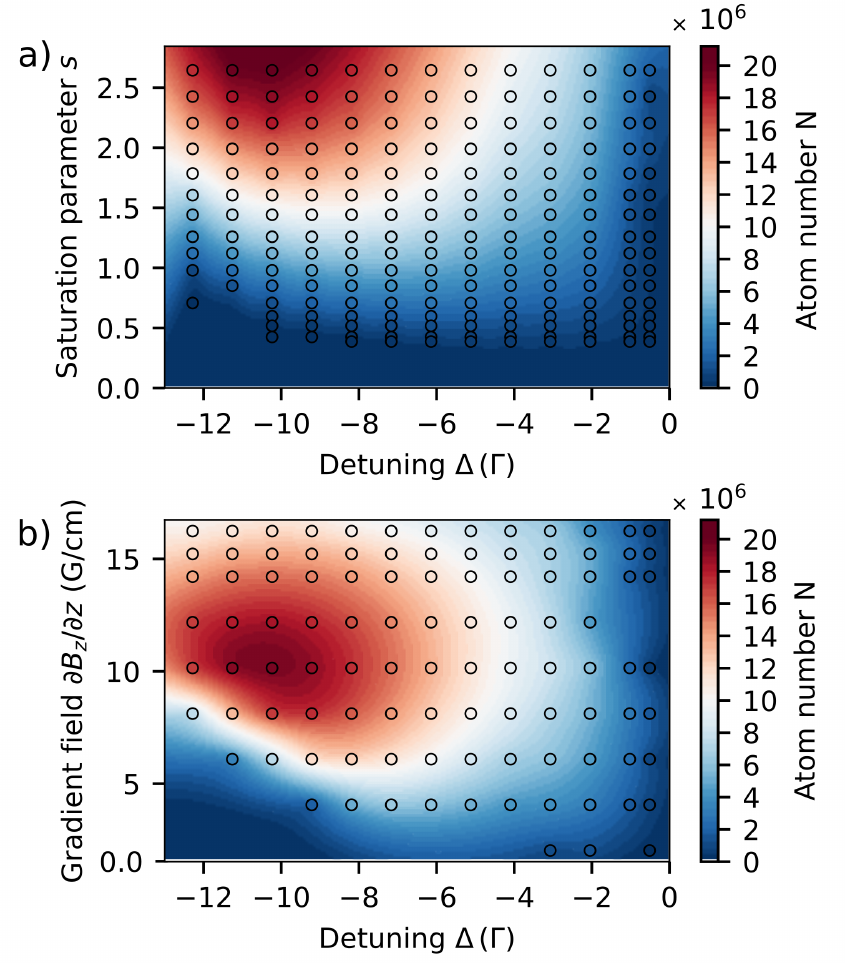}
\caption{Number of trapped atoms $N$ as a function of (a) the detuning $\Delta$ and the saturation parameter $s$ for a fixed magnetic gradient field of ${\partial B_z}/{\partial z}=10\,$G/cm, and (b) the detuning $\Delta$ and the magnetic gradient field ${\partial B_z}/{\partial z}$ for a fixed saturation parameter of $s=2.6$.}
\label{ContourMapAtomNumber}
\end{figure}

An important quantity of any cold-atom experiment is the atom number. For a fixed magnetic field gradient of ${\partial B_z}/{\partial z}= 10\,$G/cm, we investigate the dependence of atom number on the laser detuning $\Delta$ and on the saturation parameter $s=I/I_{\rm sat}$. The results are depicted in Fig.~\ref{ContourMapAtomNumber}(a). In this contour plot, the circles indicate measurement points, and the color of the circle's filling denotes the measurement value. As a background, we provide a 2D interpolation to improve the readability.

The atom number increases as the detuning increases and reaches a maximum of $2\times 10^7$ atoms around $\Delta = -10\,\Gamma$.  Beyond that maximum, the radiation pressure force becomes too weak to efficiently confine the atoms in the trap. At a detuning of $\Delta = -10\,\Gamma$, the atom number increases linearly with the saturation parameter $s$. Up to the maximum laser power available to this measurement, $s\approx 3$, we do not observe saturation of the atom number.

Fig.~\ref{ContourMapAtomNumber}(b) shows the dependence the atom number $N$ on magnetic field gradient ${\partial B_z}/{\partial z}$ and laser detuning $\Delta$ for a fixed saturation parameter $s = 2.6$. An increase of the gradient field improves the atom number until reaching a maximum around 10\,G/cm, largely independent of detuning. Beyond this maximum, a reduction of atom number is observed, explained by the reduction in capture volume at higher gradients fields. Typical atom numbers for the $^{202}{\rm Hg}$ isotope are in the range of $10^7$ atoms.

Our MOT is loaded from the background gas, and its equilibrium atom number depends on the loading rate (proportional to the Hg partial pressure) and the atom loss rate. Quite generally, the loss rate is a combination of one-particle losses (induced by collisions with room-temperature Hg atoms and all other residual gas atoms), two-particle losses (e.g.~light-assisted collisions), and three-body losses (molecule formation). For the densities obtained in this study, we conclude that only one-particle losses are relevant.

\begin{figure}
\includegraphics[width=\columnwidth, keepaspectratio]{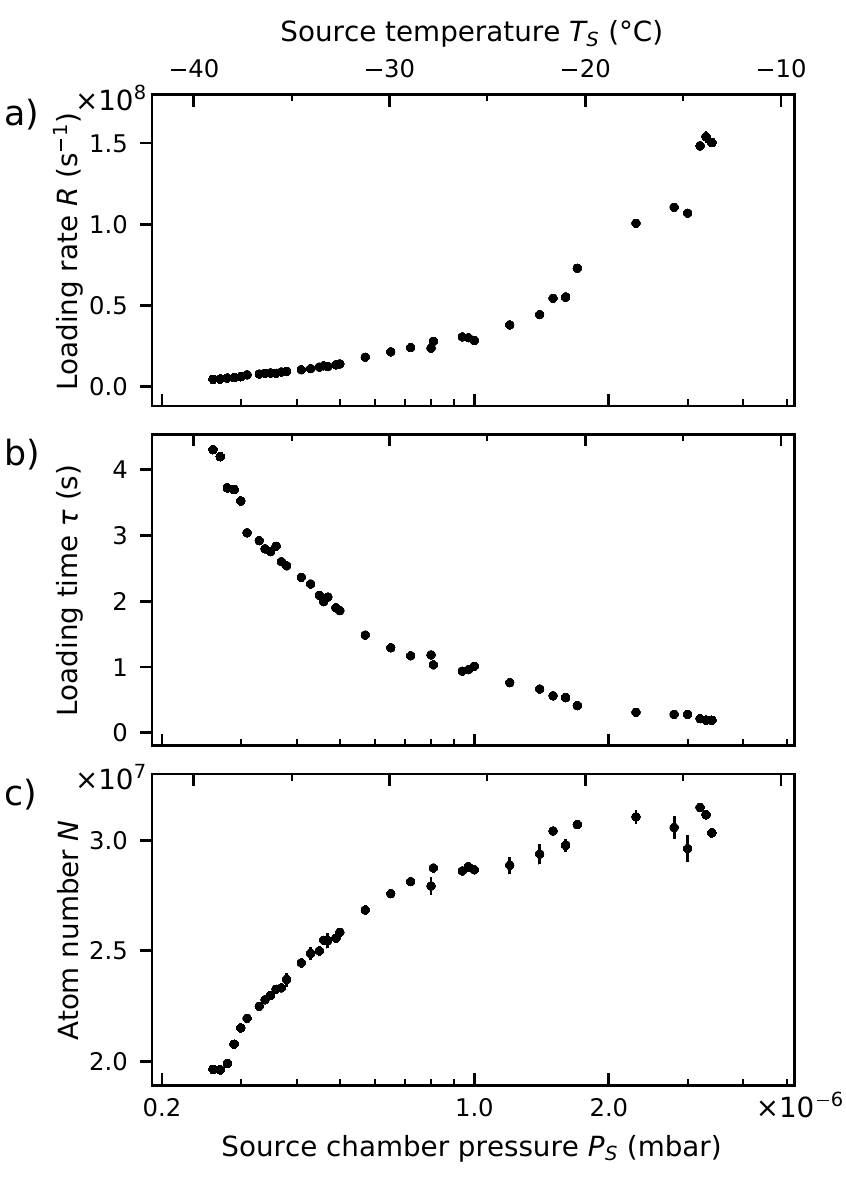}
\caption{Influence of the Hg partial pressure on (a) the initial loading rate, (b) the loading time, and (c) the equilibrium atom number. The partial pressure in the chamber can not be measured directly, shown is the pressure reading of the small ion pump in the source chamber. The top axis shows the associate source temperature.}
\label{SourceCharacteristic}
\end{figure}

We vary the Hg partial pressure through control of the oven temperature $T_s$ from $-37^{\circ}$C to $-13^{\circ}$C. The loading rate increases linearly with partial pressure; see Fig.~\ref{SourceCharacteristic}(a). The atom number saturates at a source temperature of around $-25^{\circ}$C, which corresponds to about $1.6\times10^{-6}\,$mbar in the source section. At this point, the residual gas in the vacuum chamber is dominated by mercury, and the MOT atom number becomes independent of Hg partial pressure; see Fig.~\ref{SourceCharacteristic}(c). Increasing the partial pressure further increases both the loading rate and the one-body loss rate, thus accelerates the loading dynamics, but does not increase the equilibrium atom number. A selective increase of the loading rate, and thus an increase in the MOT atom number, could be achieved through implementation of a Zeeman slower.

The maximum atom number, obtained with 35\,mW of power per MOT beam ($s=6.2$), stands at $5 \times 10^7$ atoms. We believe that even higher atom numbers could be achieved with higher laser power and a cleaner mode profile.

\subsection{Temperature}

The series of mercury isotopes lends itself well for an investigation of laser cooling mechanisms. On the one hand, the bosonic isotopes, which do not have a degenerate ground state, are particularly well suited to study simple Doppler cooling theory \cite{Lett1989}. On the other hand, the fermionic isotope $^{199}{\rm Hg}$, which has a nuclear spin of $I=1/2$, represents the simplest system which can support sub-Doppler cooling mechanisms, in particular Sisyphus cooling \cite{Dalibard1989}. The dependence of cooling performance on the number of Zeeman substates can then be explored further through the $^{201}{\rm Hg}$ isotope with nuclear spin of $I=3/2$.

\subsubsection{Dependence of the temperature on trapping parameters}

To measure the temperature $T$ of the atomic cloud, we use the time-of-flight (TOF) technique:~we release the atomic cloud from the MOT and record its ballistic expansion for a set of release times $t_{\rm TOF}$ via absorption imaging. The comparably narrow linewidth of 1.3\,MHz leads to a comparably small absorption signal. For typical temperatures of order $100\,\mu$K and atom numbers of order $10^7$, the absorption signal falls below the imaging photon shot noise at a TOF of about 10\,ms. At this point of expansion, the cloud size $r \approx \sqrt{k_BT/m}\,t_{\rm TOF}$ does not yet dominate over the initial cloud size, see Sec.~D. Therefore, we cannot assume the initial cloud size to be negligible, and each temperature measurement is obtained from a series of seven absorption images with the TOF varied between 0\,ms and 7\,ms. In this way, we can reconstruct the initial size and the expansion dynamics to infer the temperature. The radius of the cloud $r$ accessible from our two dimensional images for varying $t_\mathrm{TOF}$ corresponds to the root mean square of the fitted one-dimensional radii $r_x$ and $r_z$ along the x- and z- directions.

\begin{figure}
\includegraphics[width=\columnwidth, keepaspectratio]{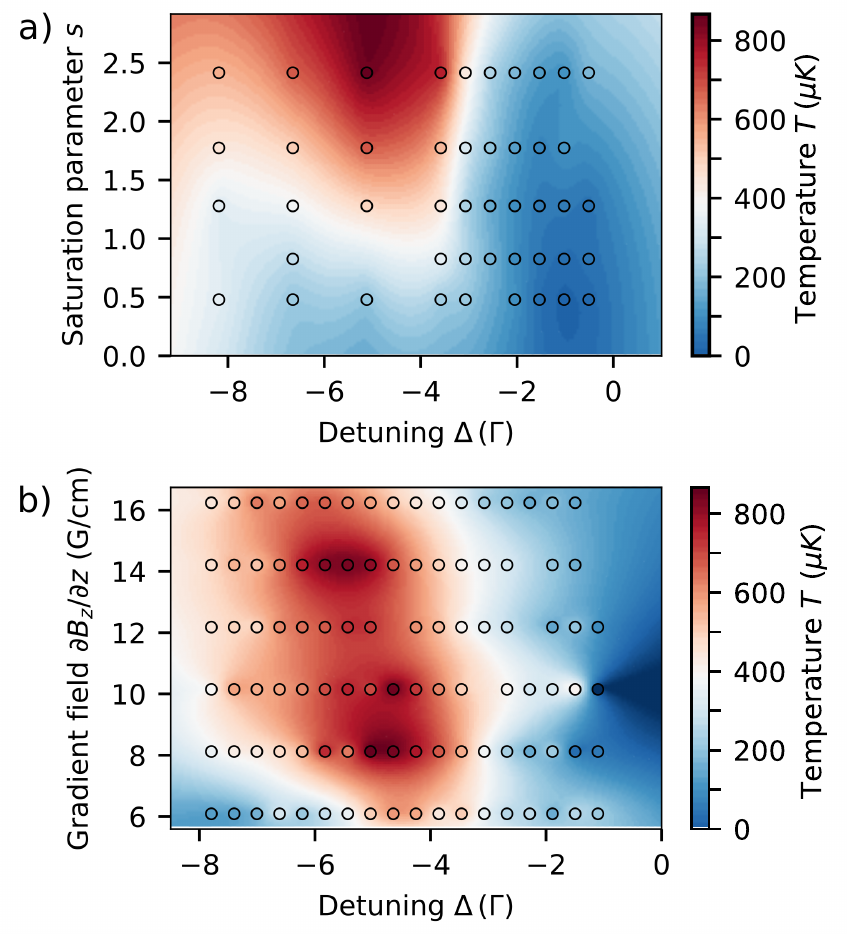}
\caption{Temperature $T$ of the cloud of atoms as a function of (a) the detuning $\Delta$ and the saturation parameter $s$ for a fixed magnetic gradient field ${\partial B_z}/{\partial z}=10 \, {\rm G/cm}$, and (b) the detuning $\Delta$ and the magnetic gradient field ${\partial B_z}/{\partial z}$ for a fixed saturation parameter $s=2.6$.}
\label{ContourMapTemperature}
\end{figure}

The dependence of the temperature $T$ on laser detuning $\Delta$ and saturation parameter $s$ is shown in Fig.~\ref{ContourMapTemperature}(a). The temperature $T$ increases with the saturation parameter $s$. Indeed, a high intensity of the MOT beams induces a heating mechanism which originates from reabsorption of scattered photons. The detuning $\Delta$ is the most critical parameter, and the lowest temperature, $T = 84(5)\,\mu$K, is obtained for a detuning of $\Delta = -\Gamma$. As shown in Fig.~\ref{ContourMapTemperature}(a), a larger detuning leads to a temperature increase of the atomic cloud.

This is also expected from one-dimensional Doppler cooling theory \cite{Lett1989}, which relates the temperature $T$ to the detuning $\Delta $ and the saturation parameter $s$,
\begin{equation}
T= \frac{\hbar \Gamma^2}{8 k_B |\Delta|} \left( 1+6s+ {\left( \frac{2 \Delta}{\Gamma} \right)}^2 \right),
\label{eqn:1}
\end{equation}
where $k_B$ is the Boltzmann constant and $\hbar$ is the reduced Planck constant.

The temperature of the cloud $T$ has been measured in function of the magnetic field gradient ${\partial B_z}/{\partial z}$ and the detuning $\Delta$ for a fixed saturation parameter $s=2.6$; see Fig.~\ref{ContourMapTemperature}(b). The gradient does not have a significant influence on the temperature $T$, as predicted by the Doppler cooling theory. In general, the temperatures observed in the experiment are higher than predicted by the Doppler cooling theory, but follow the predicted dependence on detuning and saturation parameter. This behaviour has already been observed in other experiments with alkaline-earth (-like) atoms \cite{Park2003,Xu2003,Loo2004,Kemp2016}.

\subsubsection{Sub-Doppler cooling}

We will now explore the lower limit to the temperature that can be achieved by laser cooling. As discussed above, the temperature depends only mildly on the gradient field and on the saturation parameter. Therefore, we fix these parameters to ${\partial B_z}/{\partial z} = 12.2\,$G/cm and $s = 2.7$ for the following study. For a straightforward comparison with the 1D Doppler theory, we will constrain our analysis to the temperature in the vertical (z-) direction.

\begin{figure}
\includegraphics[width=\columnwidth, keepaspectratio]{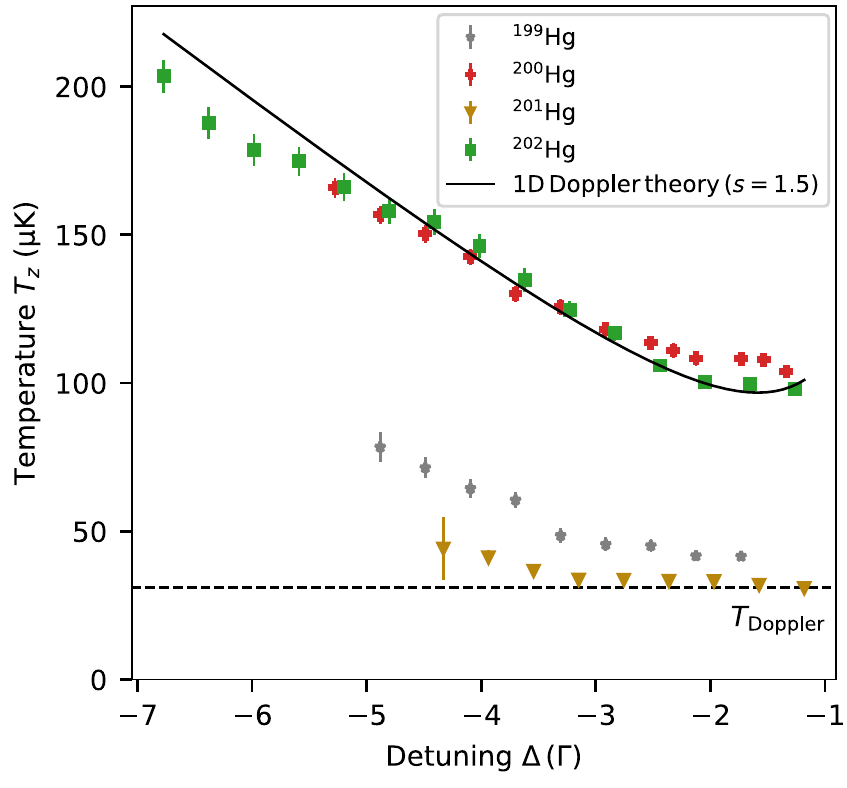}
\caption{Temperature $T$ of the atomic cloud as a function of the detuning $\Delta$, obtained for a saturation parameter of $s=2.7$ and a fixed magnetic gradient field of ${\partial B_z}/{\partial z}=12.2 \, {\rm G/cm}$ for several mercury isotopes. The Doppler limit (dashed line) is the same for both bosonic and fermionic species.}
\label{ComparisonTemperature}
\end{figure}

The temperature of the atomic cloud $T_z$ as a function of detuning $\Delta$ for two bosonic and two fermionic isotopes is presented in Fig.~\ref{ComparisonTemperature}. Each data point is the weighted average of at least five time-of-flight sequences. Each sequence is composed of 0.5-ms steps and lasts until the disappearance of the signal. The temperature of the bosonic species $^{202}{\rm Hg}$ ($^{200}{\rm Hg}$) reaches a minimum at 98(2)\,$\mu$K (104(3)\,$\mu$K) at $\Delta= -1.6 \, \Gamma$ ($\Delta= -1.5 \, \Gamma$).

To compare now our results with the Doppler theory, we fit our data with the expression from Eq.~(\ref{eqn:1}), where we leave the saturation $s$ as a free parameter. The model fits the measured temperatures well, but the derived saturation parameters are slightly lower ($s=1.5(1)$ for $^{200}{\rm Hg}$ and $s=1.8(1)$ for $^{202}{\rm Hg}$) compared to the experimentally measured intensities. This difference is caused by the non-Gaussian profile of the MOT beams:~when determining the peak intensity of the beams, and thus the saturation parameter, we assume the beam shape to be Gaussian, and systematically overestimate the peak intensity for non-Gaussian beam profiles.

In summary, we confirm that the cooling mechanism of bosonic mercury isotopes is properly described by Doppler theory \cite{Mcferran2010}. The lack of degenerate ground states ($I=0$) precludes sub-Doppler cooling mechanisms. This situation is different for the fermionic isotopes $^{199}{\rm Hg}$ and $^{201}{\rm Hg}$, which do possess multiple Zeeman substates and indeed show temperatures substantially lower than their bosonic counterparts.

The cloud of $^{199}{\rm Hg}$ atoms has a temperature $\sim 40\, \mu{\rm K}$ for a detuning between $2\, \Gamma$ and $3\, \Gamma$. Above $3\,\Gamma$, the temperature increases. The $^{201}{\rm Hg}$ atoms reach the lowest temperature of $30.9(2.3)\,\mu{\rm K}$, right at the Doppler temperature $T_D=31 \,\mu{\rm K}$. These two fermionic species undergo Sisyphus cooling, but there is a subtle difference in the number of Zeeman substates. Indeed, ground-state level degeneracy is the key parameter of sub-Doppler cooling because it affects the velocity capture range \cite{Xu2003}. Thus, the richer atomic structure of the $^{201}{\rm Hg}$ is an asset to reach lower temperatures than $^{199}{\rm Hg}$. Mercury appears to be a promising system to study the interplay between Doppler and sub-Doppler cooling mechanisms \cite{Chang2014,Yudkin2018}.

\subsection{Cloud size and atomic density}

\subsubsection{Cloud size and Doppler theory}

\begin{figure}
\includegraphics[width=\columnwidth, keepaspectratio]{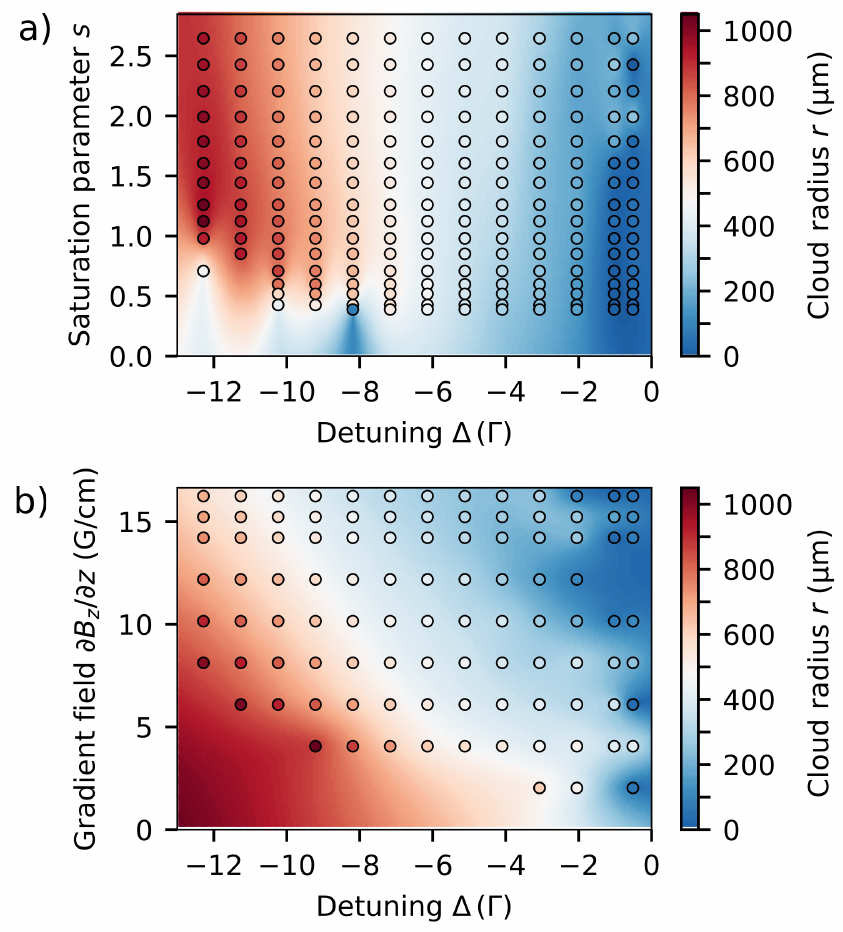}
\caption{Radius $r$ of the atomic cloud as a function of (a) the detuning $\Delta$ and the saturation parameter $s$ for a fixed magnetic gradient field ${\partial B_z}/{\partial z}=10 \, {\rm G/cm}$, and (b) the detuning $\Delta$ and the magnetic gradient field ${\partial B_z}/{\partial z}$ for a fixed saturation parameter $s=2.6$. }
\label{ContourMapRadiusCloud}
\end{figure}

Cloud radius is an important parameters when studying the performance of a MOT. It directly reflects the spatial arrangement and its evolution the complex dynamic of atoms within the cloud. From the same measurements used to generate Fig.~\ref{ContourMapAtomNumber}(a), we extract the radius $r = \sqrt{{r_x}^2 + {r_z}^2}$ of the atomic cloud in function of detuning $\Delta$ and saturation parameter $s$; see Fig.~\ref{ContourMapRadiusCloud}.
We observe, that the cloud size increases with detuning, it decreases with magnetic field gradient, and is rather independent of the light intensity.

\begin{figure}
\includegraphics[width=8.6 cm, keepaspectratio]{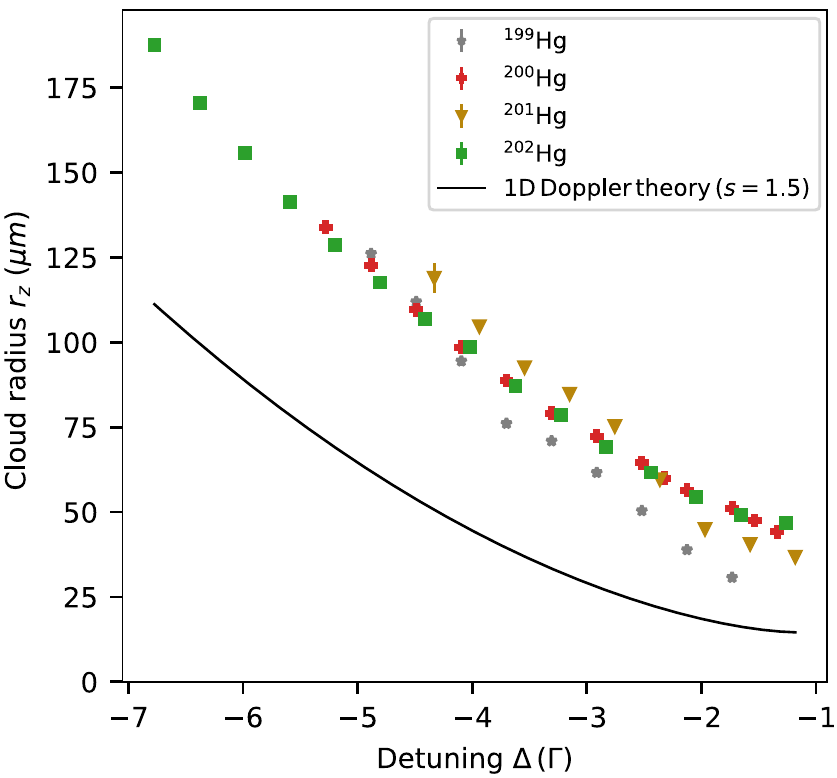}
\caption{Radius $r_z$ of the cloud of atoms as a function of the detuning $\Delta$ with $s=2.7$ and ${\partial B_z}/{\partial z}=12.2 \, {\rm G/cm}$ for several mercury isotopes. The black line shows the Doppler theory prediction, assuming $s=1.5$ as determined from the temperature measurement.}
\label{ComparisonMapRadiusCloud}
\end{figure}

Considering the good agreement of the Doppler theory for the bosonic species, we will compare the predicted radius of the cloud with our data in the z direction, see Fig.~\ref{ComparisonMapRadiusCloud}. Using the equipartition theorem, the radius $r$ and temperature $T$ of the cloud are related through
\begin{equation}
\frac{1}{2} \kappa r^2=\frac{1}{2} k_B T.
\label{eqn:2}
\end{equation}
Here, $k_B$ is the Boltzmann and $\kappa$ is the trap spring constant, which can be expressed as
\begin{equation}
\kappa= \frac{8 k \Delta}{\Gamma} \frac{6s}{(1+6s+(\frac{2\Delta}{\Gamma})^2)^2} g_j \mu_B \frac{\partial B_z}{\partial z},
\label{eqn:3}
\end{equation}
where $\mu_B$ is the Bohr magneton, $g_j$ the Landé factor of the excited state, and $k$ the photon wavevector \cite{Lett1989}.

Combining Eqs.~\ref{eqn:1}, \ref{eqn:2}, and \ref{eqn:3}, we can obtain an expression for the radius of the cloud:
\begin{equation}
r_z= \sqrt{\frac{\hbar \Gamma^3}{64 \Delta^2 k \, g_j \mu_B} \frac{(1+6s+(\frac{2\Delta}{\Gamma})^2)^3}{6s} \left( {\frac{\partial B_z}{\partial z}}\right)^{-1}}.
\label{eqn:4}
\end{equation}

The dependence of cloud size on detuning is shown in Fig.~\ref{ComparisonMapRadiusCloud}: the cloud radius grows with the detuning. Moreover, the size of the cloud is largely independent of atom number. The clouds are roughly two times larger than predicted by theory if we use the value of the saturation parameter $s=1.5$, as derived from the temperature measurement in Fig.~\ref{ComparisonTemperature}. Using the experimentally determined saturation parameter of $s=2.7$ provides around $10\%$ lower predicted temperature than the real temperature. Related studies have observed larger-than-expected cloud sizes as well \cite{Mcferran2010}. The discrepancy is likely explained by inhomogeneous and non-Gaussian beam profiles, as well as the effective repulsion between atoms from absorption/re-emission cycles of cooling light.

\subsubsection{Atomic density}

\begin{figure}
\includegraphics[width=\columnwidth, keepaspectratio]{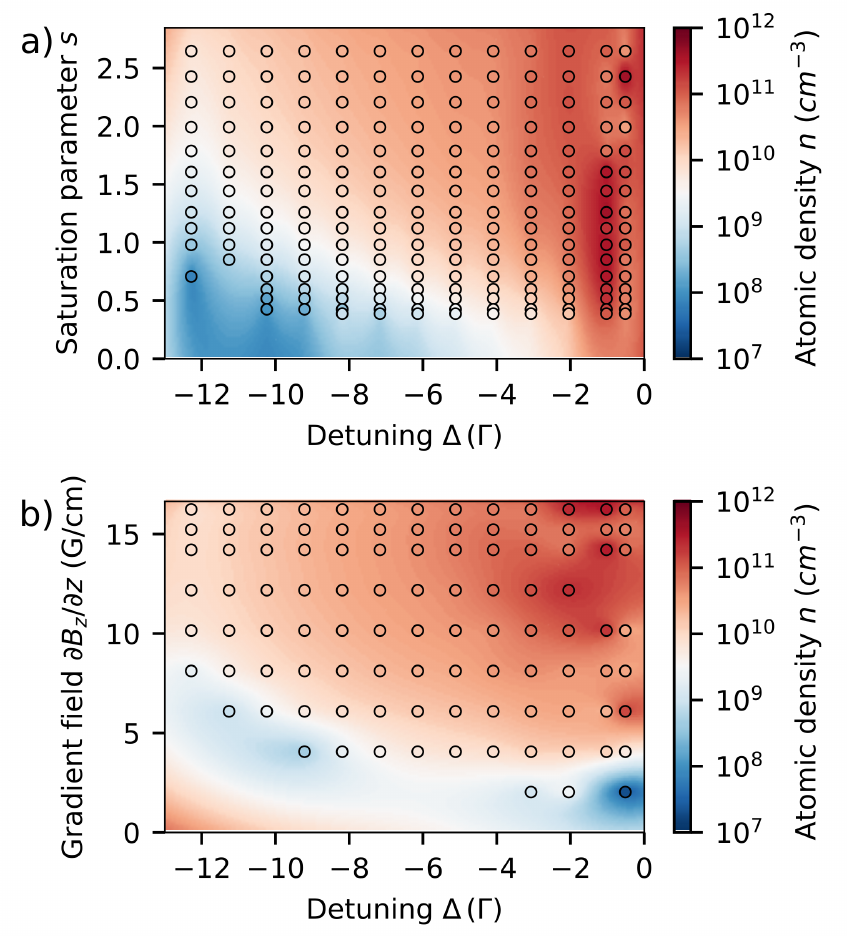}
\caption{Atomic density $n$ of the atomic cloud as a function of (a) the detuning $\Delta$ and the saturation parameter $s$ for a fixed magnetic gradient field ${\partial B_z}/{\partial z} = 10\,$G/cm, and (b) the detuning $\Delta$ and the magnetic gradient field ${\partial B_z}/{\partial z}$ for a fixed saturation parameter $s = 2.6$.}
\label{ContourMapAtomicDensityCloud}
\end{figure}

The cloud of atoms is considered to homogeneously occupy a volume $V=\frac{4}{3} \pi r_x\, r_y\, r_z$, where $r_i$ is the cloud radius in a given direction $i$. The atomic density then follows as $n=N/V$. To maximize the densiy, we identify an optimum detuning near $\Delta=-\Gamma$; see Fig.~\ref{ContourMapAtomicDensityCloud}(a). The density favors large gradient fields and mild saturation parameters. The highest densities of the bosonic isotope $^{202}{\rm Hg}$ are observed for a gradient field between 10\,G/cm and 15\,G/cm and reach a value of $n_{^{202}} = 4.1 \times 10^{11} \, {\rm{cm}}^{-3}$. Increased loss mechanisms, such as light-assisted inelastic collisions \cite{Julienne1992}, as well as photon re-absorption \cite{Sesko1991} lead to a saturation of the density for even higher gradient fields.

\subsection{Phase space density}

The phase space density $\rho$ is the relevant quantity in the context of degenerate quantum gases \cite{Townsend1995}. It is expressed as
\begin{equation}
\rho= n \Lambda ^3 = n \frac{\hbar \sqrt{2 \pi}}{\sqrt{m k_B T}},
\label{eqn:5}
\end{equation}
where $k_B$ is the Boltzmann constant, $\hbar$ the reduced Planck constant, and $m$ the mass of an atom.

\begin{figure}
\includegraphics[width=\columnwidth, keepaspectratio]{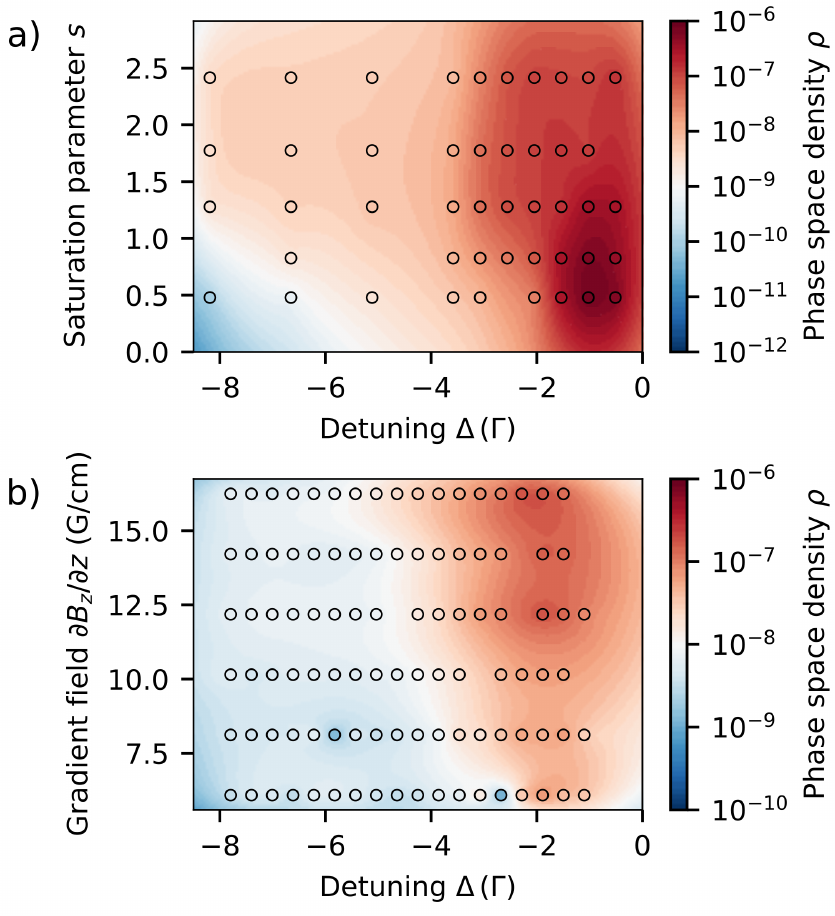}
\caption{Phase space density $\rho $ of $^{202}{\rm Hg}$ as a function of (a) detuning $\Delta$ and saturation parameter $s$ for a fixed magnetic field gradient ${\partial B_z}/{\partial z}=10\,$G/cm, and (b) as a function of detuning $\Delta$ and magnetic gradient field ${\partial B_z}/{\partial z}$ at a fixed saturation parameter of $s=2.6$.}
\label{ContourMapPhaseSpacedensity}
\end{figure}

The dependence of the phase space density $\rho$ on the detuning $\Delta$ and the saturation parameter $s$ is shown in Fig.~\ref{ContourMapPhaseSpacedensity}(a). The highest phase space density is obtained for low saturation parameters $s \leq 1$, which avoids heating of the cloud through reabsorption of scattered photons. In terms of detuning, adjusting the frequency of the laser close to resonance is beneficial to minimize the cloud temperature. Thus, favoring cooling over atom number is the best strategy to maximize the phase space density. In our experiment, a detuning of $\Delta \approx -\Gamma$ provides the highest phase space density.

Moreover, the phase space density grows with the gradient of the magnetic field; see Fig.~\ref{ContourMapPhaseSpacedensity}(b), as the trap volume is reduced. The highest phase space density for the bosonic isotope $^{202}{\rm Hg}$ is $\rho_{202} = 7.5(7) \times 10^{-7}$, reached at a gradient field of $ {\partial B_z}/{\partial z}= 12.5\,$G/cm. For higher gradient fields, we expect that the scattering losses increase and reduce the phase space density as suggested by the  parameters to obtain the highest atom number $N$ on Fig.~\ref{ContourMapAtomNumber}.

For the fermionic isotopes $^{199}{\rm Hg}$ and $^{201}{\rm Hg}$, we also perform a measurement of the phase space density as a function of detuning $\Delta$ and saturation $s$ at a gradient of ${\partial B_z}/{\partial z}=12.2\,$G/cm. The picture resembles the bosonic case:~highest phase space densities are obtained for small detuning and small intensity. Specifically, we obtain $\rho_{199} = 1.9(2) \times 10^{-6}$ and $\rho_{^{202}{\rm Hg}} =  6.1(6) \times 10^{-6}$.

These numbers provide a very promising basis for dipole trap loading to further increase the phase space density. Here, dynamic compression and cooling phases could be implemented. Evaporative cooling, \textit{en route} to quantum degeneracy, will increase the phase space density further.

\section{Conclusion}
In conclusion, we have presented an in-depth study laser cooling of mercury. With more laser power than available in previous experiments, we scanned the three-dimensional parameter space of laser detuning, field gradient, and laser intensity. An optimum set of parameters allowed us to increase the number of trapped atoms by about an order of magnitude compared to previous studies. Inhomogeneities in the laser's mode profile reduce the cooling performance and lead to a discrepancy between the calculated and measured temperature and MOT size in dependence of laser intensity.
We show that sub-Doppler cooling for the two fermionic isotopes closely follows theoretical expectations. We obtain phase space densities of order $10^{-6}$, which appear to be a solid basis for dipole trap loading. It is interesting to note that the phase space density obtained with the fermionic isotopes is about an order of magnitude larger than for the bosonic counterparts: clearly, the sub-Doppler cooling mechanisms overcompensate the smaller capture efficiency. Work towards quantum degeneracy would benefit from the implementation of a Zeeman slower to reduce the background pressure and improve the loading rate.

\begin{acknowledgments}
We acknowledge fruitful discussions with A.~Yamaguchi, A.~Widera, M.~Köhl, J.~Kroha, and D.~Meschede, and we thank F.~Affeld for assistance in the operation of the experiment. Financial support from the DFG through SFB TR 185 ``OSCAR'', project number 277625399, as well as from the ERC, project number 757386  ``quMercury", is gratefully acknowledged.
\end{acknowledgments}

\bibliographystyle{apsrev}
\bibliography{bib}

\end{document}